\documentclass[aps,amsfonts,amsmath,prd,preprint,nofootinbib]{revtex4-1}
\pdfoutput=1
\usepackage{graphicx}

\newcommand{\beq}{\begin{equation}}
\newcommand{\eeq}{\end{equation}}
\newcommand{\bea}{\begin{eqnarray}}
\newcommand{\eea}{\end{eqnarray}}

\def\({\left(}
\def\){\right)}
\def\[{\left[}
\def\]{\right]}
\def\a{\alpha}

\begin{document}

\title{Tunneling decay rate in quantum cosmology}
\author{Audrey T. Mithani$^{1,2}$, Alexander Vilenkin$^1$}
\address{$^1$Institute of Cosmology, Department of Physics and Astronomy,
\\
Tufts University, Medford, MA 02155, USA \\
$^2$Center for Cosmology and Particle Physics, Department of Physics, \\
New York University, New York, NY 10003, USA}

\begin{abstract}
In canonical quantum cosmology, the wave function of the universe lacks explicit time dependence.  However, time evolution may be present implicitly through the semiclassical superspace variables, which themselves depend on time in classical dynamics.  In this paper, we apply this approach to an oscillating universe model recently introduced by Graham et 
al.  By extending the model to include a massless, minimally coupled scalar field $\phi$ which has little effect on the dynamics but can play the role of a ``clock'', we determine the decay rate of the oscillating universe.
\end{abstract}
\maketitle

\section{Introduction}

Oscillating cosmological models have been extensively studied over the years (see, e.g., \cite{Dabrowski95,Graham11} and references therein).  Such models can be classically stable with respect to small perturbations \cite{Graham11,MV14,Graham14}, but it was pointed out in \cite{DL95b,MV12} that they are generically unstable with respect to decay through quantum tunneling to zero size.  Our goal in the present paper is to calculate the corresponding decay rate.

In the semiclassical regime, the decay rate can be expressed as
\beq
\Gamma={\cal A} e^{-2|S_E|}, 
\eeq
where $S_E$ is the under-barrier Euclidean action.  The action $S_E$ has been calculated in Refs.~\cite{DL95b,MV12} for some simple FRW models.  Here, we would like to go beyond that and also calculate the pre-exponential factor ${\cal A}$, at least in the framework of the FRW models under consideration.

The problem we have to address is that the rate $\Gamma$ is the decay probability per unit time, and the time variable is conspicuously absent in the formalism of quantum cosmology.  In particular, the wave function of the universe $\Psi$ is independent of time.  The way out of this impasse was already pointed out by DeWitt \cite{DeWitt}, who suggested that the role of a clock could be played by some geometric or matter variable.   Any time evolution should then be understood implicitly, in terms of the canonical variables themselves.  We shall adopt this approach here and use it to calculate the decay rate in the ``simple harmonic universe'' (SHU) model of Ref.~\cite{Graham11}, suitably extended to include a clock.

The SHU is a closed universe with energy density due to a negative cosmological constant $\Lambda <0$ and a matter component having equation of state $P=w\rho$.  Oscillating solutions exist for $-1 < w <-1/3$.  The gravity of matter with such equation of state is repulsive; it causes a contracting universe to bounce and to start expanding.  On the other hand, a negative $\Lambda$ causes an expanding universe to turn around and start contracting.  As a result, the universe will oscillate between some maximum and minimum size.  Here, we focus on the case where $w = -2/3$ for calculation simplicity, though the situation is qualitatively similar for other values of $w$.

The FRW model of Ref.~\cite{Graham11} includes a single dynamical variable --  the scale factor $a$.  In the case of an oscillating universe, the scale factor evolution is not monotonic; hence it cannot serve as a time variable.  We therefore introduce a second minisuperspace variable -- a homogeneous, massless, minimally coupled scalar field $\phi$, which will play the role of a clock.  We shall assume that the contribution of this field to the total energy density of the SHU is negligible, so that its presence has little effect on the dynamics of oscillations and does not alter the stability analysis of \cite{Graham14}. 

The rest of the paper is organized as follows.  In the next section we review the SHU model and extend it by introducing a homogeneous scalar field.  The semiclassical wave function for this model is found in Sections III and IV.  The tunneling rate and the corresponding lifetime of the universe are calculated in Sec. V.  This rate has the expected relation to the tunneling probability, as we show in Sec. VI.  Finally, in Sec. VII we give a brief summary and some concluding remarks.

\section{Simple harmonic universe with a clock}

\subsection{Simple harmonic universe}

Classical dynamics of the SHU are described by the $k=+1$ Friedmann equation
\beq
\frac{\dot{a}^2}{a^2} + \frac{1}{a^2} = \frac{8\pi G}{3} \rho(a),
\label{Friedmann}
\eeq
with energy density
\beq
\rho(a)=\Lambda + \frac{\sigma}{a}.
\label{rhoa}
\eeq
Here, $\Lambda < 0$ is the cosmological constant and the second term in (\ref{rhoa}) describes ``domain wall matter'' with equation of state $w=-2/3$.  We shall assume that both parameters $\Lambda$ and $\sigma$ are small in Planck units,
\beq
|\Lambda|\ll G^{-2},  ~~~ \sigma \ll G^{-3/2}.
\label{small}
\eeq

The Friedmann equation has oscillating solutions for the scale factor
\beq
a(t) = \omega^{-1}(\gamma - \sqrt{\gamma^2 - 1}\cos(\omega t)),
\eeq
where
\beq
\omega = \sqrt{\frac{8\pi G}{3}|\Lambda|} \quad ; \quad \gamma = \sqrt{\frac{2\pi G}{3}\frac{\sigma^2}{|\Lambda|}}
\eeq
and we have to require that $\gamma > 1$.  The universe oscillates between maximum and minimum values of
\beq
a_{\pm} = \omega^{-1}\(\gamma \pm \sqrt{\gamma^2 - 1}\).
\eeq

Alternatively, the system also may be described by the constrained Hamiltonian
\beq
 {\cal H} = -\frac{G}{3\pi a} \(p_a^2 -\tilde U(a)\) = 0,
\eeq
where the momentum conjugate to $a$ is
\beq
p_a = -\frac{3\pi}{2G} a \dot a
\eeq
and the potential $\tilde U(a)$ is
\beq
\tilde U(a) = \( \frac{3\pi}{2G} \)^2 a^2 \( 1-\frac{8\pi G}{3} a^2 \rho(a) \).
\label{U}
\eeq

In the context of quantum cosmology, the momentum becomes the differential operator $p_a \to -i\partial / \partial a$ and the wave function of the universe $\Psi(a)$ satisfies the Wheeler-deWitt equation
\beq
{\cal {H}} \Psi (a) = 0.
\eeq

\subsection{SHU with a scalar field}

We now modify the SHU model by adding a homogeneous, massless, minimally coupled scalar field $\varphi(t)$.  The corresponding Hamiltonian constraint is
\beq
{\cal H} = -\frac{G}{3\pi a} \(p_a^2 - \frac{3}{4\pi G a^2}p_{\varphi}^2 +\tilde U(a)\) = 0,
\eeq
where
\beq
p_{\varphi} =  2\pi^2 a^3 \dot \varphi
\label{pphi}
\eeq
is the momentum conjugate to $\phi$.  The momentum $p_\phi$ is a constant of motion, $\dot p_\varphi = 0$.  Without loss of generality, we shall assume that $p_\varphi >0$.  Then it follows from Eq.~(\ref{pphi}) that the scalar field $\varphi$ increases monotonically; hence it can be used as a time variable.

In quantum cosmology, we make the replacement $p_a \to -i\partial / \partial a$ and $p_\varphi \to -i \partial / \partial \varphi$, and the Hamiltonian constraint ${\cal H} = 0$ becomes the Wheeler DeWitt (WDW) equation
\beq
\[ -a\frac{\partial}{\partial a}a \frac{\partial}{\partial a} + a^2 \tilde U(a) + \frac{3}{4\pi G}\frac{\partial^2}{\partial \varphi^2} \] \Psi(a,\phi) = 0.
\label{WDW1}
\eeq
Here, we have adopted the ordering of the non-commuting factors $a$ and $\partial/\partial a$ proposed in \cite{Misner}, for which the differential operator in Eq.~(\ref{WDW1}) becomes a covariant Laplacian.  

With the change of variables $\alpha = \ln \(\omega \gamma a\)$, $\phi = (4\pi G/3)^{1/2} \varphi$,
the WDW equation becomes
\beq
\[ -\frac{\partial^2}{\partial\alpha^2} + U(\alpha) + \frac{\partial^2}{\partial \phi^2}\] \Psi(\alpha,\phi) = 0,
\label{WDW2}
\eeq
where the potential $a^2 \tilde U(a) =  U(\alpha)$ is
\beq
U(\a) = \beta^{-2}e^{4\a}\( 1-2 e^{\a} + \gamma^{-2} e^{2\alpha} \),
\eeq
where 
\beq
\beta = \(\frac{2G}{3\pi}\)\omega^{2}\gamma^2 = \frac{32\pi}{27}G^3\sigma^2.
\label{beta}
\eeq

The WDW equation (\ref{WDW2}) separates, and the general solution can be expresses as a superposition of terms of the form
\beq
\Psi(\alpha,\phi) = e^{i p\phi} f_{p}(\alpha),
\label{solns}
\eeq
where the separation parameter $p$ is the eigenvalue of the momentum $p_\phi$ and the function $f_p(\alpha)$ satisfies the equation
\beq
\[ -\frac{\partial^2}{\partial \alpha^2} + U_p(\alpha)\] f_p(\alpha) = 0
\label{WDW}
\eeq
with
\beq
U_p(\a)=U(\a)-p^2.
\eeq
The effective potential $U_p(\a)$ is plotted in Fig.~1.

We see that inclusion of a scalar field has the effect of decreasing the potential $U(\alpha)$ by a constant term, $-p^2$.  We assume that this term is small compared to the characteristic scale of the potential, that is,
\beq
p\ll\beta^{-1}.
\label{pbeta}
\eeq
This term, however, does have an effect near the turning points 
\beq
\a_\pm = \ln\( \gamma^2 \pm \gamma^2 \sqrt{1-\gamma^{-2}} \)
\eeq
of the unperturbed potential, $U(\a_\pm)=0$.  The turning points in the presence of a scalar field,
\bea
\a_1 &=& \a_- - \delta \a_1 \\
\a_2 &=& \a_+ + \delta \a_2
\label{tp1}
\eea
can be found by solving  $U_p(\a) = 0$.
To the lowest order in $p^2$, we have
\bea
\delta \a_1 &\simeq& \frac{p^2}{|U'(\a_-)|} 
\label{deltaalpha1} \\
\delta \a_2 &\simeq& \frac{p^2}{U'(\a_+)}.
\label{deltaalpha2}
\eea
The derivatives of the potential appearing in Eqs.~(\ref{deltaalpha1}),(\ref{deltaalpha2}) are
\bea
U'(\a_-) & = & -\frac{1}{\beta^2} \( -1+\sqrt{1-\gamma^{-2}} \)^4 \gamma^8\( 1+\( -1+\sqrt{1-\gamma^{-2}} \)\gamma^2 \), \\
U'(\a_+) & = & \frac{1}{\beta^2} \( 1+\sqrt{1-\gamma^{-2}} \)^4 \gamma^8\( -1+\( 1+\sqrt{1-\gamma^{-2}} \)\gamma^2 \) ,
\eea
or, by order of magnitude,
\beq
U'(\a_-)\sim -\beta^{-2}, ~~~ U'(\a_+)\sim \beta^{-2}\gamma^{10}.
\label{U'}
\eeq
Since $\gamma\gtrsim 1$, we can write
\beq
\delta \a_\pm \lesssim \beta^2 p^2 \ll 1.
\eeq

Apart from shifting the turning points $\a_\pm$, the scalar field also modifies the character of the potential at small $a$ ($\alpha\to -\infty$), introducing another classically allowed region (region I in Fig.~1).  The potential at $\alpha\to -\infty$ can be approximated as $U_p(\a) \sim \beta^{-2} e^{4\a} - p^2$, so the boundary of this region is approximately
\beq
\a_0 \simeq \frac{1}{2}\ln (\beta p) .
\label{tp0}
\eeq
In order to justify semiclassical treatment, we shall require that the corresponding scale factor is large in Planck units,
\beq
a_0 =\left(\frac{2Gp}{3\pi}\right)^{1/2} \gg G^{1/2} ,
\eeq
which implies $p\gg 1$.  

\begin{figure}
\includegraphics[width=0.8\textwidth]{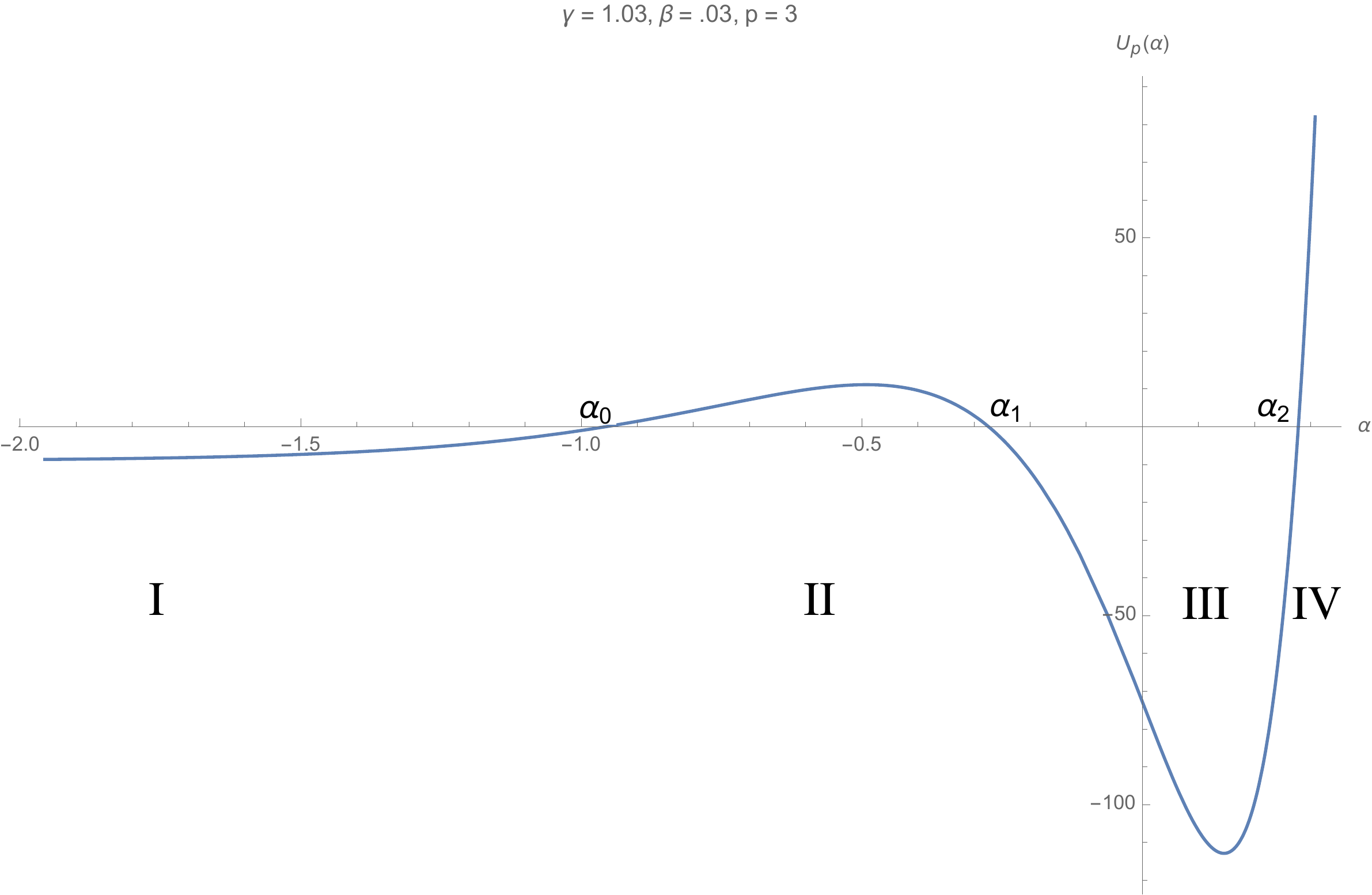}
\caption{Effective potential $U_p(\a)$, with $\gamma = 1.03$, $\beta = .03$, and $p=3$.}
\end{figure}

The two classically allowed regions are separated by a barrier (region II) extending between the turning points $\alpha_0$ and $\a_1$.  The situation is therefore analogous to a particle in a metastable state: the particle is localized in an approximate energy eigenstate, but the corresponding energy eigenvalue takes on an imaginary part, indicating a non-vanishing decay rate.  We shall see that the wave function in our model exhibits a very similar behavior.

\section{Semiclassical solutions}

Solutions to the WDW equation are specified by imposing boundary conditions appropriate to the problem.  Here, we first require that the wave function must decay under the infinite barrier to the right of $\a_2$, $f_p(\a) \to 0$ as $\a \to \infty$ (this is the same boundary condition chosen in \cite{MV12}).  In addition, we require that the solution be outgoing (left-moving) in the region $\a<\a_0$.  Physically, this means that once the oscillating universe tunnels through the barrier, it collapses to the singularity at $a=0$ $(\alpha\to -\infty)$.  In other words, the singularity is a point of no return: the probability of getting back from $a=0$ to $a=a_0$ is zero.

Sufficiently far from the turning points, we can determine the solutions using the semiclassical approximation:
\beq
f_p(\a) \simeq \frac{C_1 e^{-i\pi/4}}{[-U_p(\a)]^{1/4}} e^{+ i\int_{\a_*}^{\a}\sqrt{-U_p(\a')} d\a'} + \frac{C_2 e^{i\pi/4}}{[-U_p(\a)]^{1/4}} e^{- i\int_{\a_*}^{\a}\sqrt{-U_p(\a')} d\a'}.
\eeq
for left ($+$) and right ($-$) moving waves in the classically allowed region, $U_p(\a) < 0$, and
\beq
f_p(\a) \simeq \frac{D_1}{(U_p(\a))^{1/4}} e^{+ \int_{\a_*}^{\a}\sqrt{U_p(\a')} d\a'} + \frac{D_2}{(U_p(\a))^{1/4}} e^{- \int_{\a_*}^{\a}\sqrt{U_p(\a')} d\a'}.
\eeq
for growing ($+$) and decaying ($-$) solutions under the barrier, $U_p(\a)>0$.

Near a turning point, $\alpha=\a_*$, where\footnote{We have to use the real part, since the parameter $p$ is generally complex, and thus the potential $U_p(\a)$ is also complex.  We shall see, however, that the imaginary part of $p$ is exponentially suppressed.  Hence we shall disregard it everywhere, except for the calculation of the decay rate.} 
\beq
{\rm {Re}}~ U_p(\a_*)=0, 
\eeq
the semiclassical approximation breaks down.  In such regions we use the standard technique \cite{Landau} and approximate the potential by a linear function,
\beq
U(\a_* + \delta \a) \simeq U(\a_*) + U'(\a_*) \delta \a
= U'(\a_*) (\a -\a_*).
\eeq
Setting $z = (U'(\a_*))^{1/3} (\a - \a_*)$, the approximate WDW equation near a turning point is
\beq
\( \frac{\partial^2}{\partial z^2} - z \) \Psi(z) = 0.
\eeq
The solution is a linear combination of Airy functions $Ai(z)$ and $Bi(z)$, having the asymptotic ($z \to \infty$) forms
\bea
Ai(z) &\sim& \frac{1}{2\sqrt{\pi}} z^{-1/4} e^{-\frac{2}{3}z^{3/2}} \\
Bi(z) &\sim& \frac{1}{\sqrt{\pi}} z^{-1/4} e^{\frac{2}{3}z^{3/2}} \\
Ai(-z) &\sim& \frac{1}{\sqrt{\pi}} z^{-1/4} \sin \[ {\frac{2}{3}z^{3/2}}  +\frac{\pi}{4} \] \\
Bi(-z) &\sim& \frac{1}{\sqrt{\pi}} z^{-1/4} \cos \[{\frac{2}{3}z^{3/2}} +\frac{\pi}{4} \].
\eea

On the other hand, for the linearized potential,
\bea
\int_{\a_*}^\a [U(\a)]^{1/2} d\a \simeq [U'(\a_{*})]^{1/2}  \int_{\a_*}^\a (\a - \a_*)^{1/2} d\a &=& \frac{2 [U'(\a_{*})]^{1/2}}{3}(\a - \a_*)^{3/2} \\
&=& \frac{2}{3}z^{3/2}.
\eea
We thus see that the linearized approximation near the turning points, with Airy function solutions, matches onto the WKB solutions away from the turning points.  We determine solutions in all regions by imposing the boundary conditions in regions I and IV, and match semiclassical solutions across the turning points $\alpha_0$, $\a_1$ and $\a_2$..

We first apply the boundary condition in region IV, to the right of $\alpha_2$. There, the solution consists only of a decaying mode, $f_p(\alpha\to\infty) \to 0$:
\beq
f_p^{IV}(\a) = \frac{A}{2(U_p(\a))^{1/4}} e^{-\int^{\a}_{\a_2}\sqrt{U_p(\a')} d\a'},
\eeq
where $A={\rm const}$.  With the asymptotic form of the Airy functions, this fixes the coefficients across $\a_2$ in region $III$:
\beq
f_p^{III}(\alpha) = \frac{e^{-i\pi/4}A}{2[-U_p(\alpha)]^{1/4}} \( e^{i\int_{\alpha}^{\alpha_2}\sqrt{-U_p(\a')} d\a'} + ie^{-i\int_{\a}^{\a_2}\sqrt{-U_p(\a')} d\a'} \).
\label{f3a}
\eeq

The second boundary condition -- that the solution be outgoing in the region $\a < \a_0$ -- means that the solution must take the form
\beq
f_p^{I}(\alpha) = \frac{e^{-i\pi/4}}{[-U_p(\a)]^{1/4}} B e^{-i\int_{\a}^{\a_0}\sqrt{-U_p(\a')} d\a'},
\eeq
where $B$ is a constant coefficient.\footnote{Note that two linearly independent solutions in the limit $\alpha\to -\infty$ are $f_p(\a)\propto \exp(\pm ip\a)$, so the outgoing mode can be unambiguously identified.}  
We can now use the same method as above to match the solutions across the turning point $\a_0$ and fix the coefficients of the wave function in region II:
\beq
f_p^{II}(\a) = \frac{-iB}{2[U_p(\a)]^{1/4}} e^{-\int^{\a}_{\a_0}\sqrt{U_p(\a')} d\a'} + \frac{B}{[U_p(\a)]^{1/4}} e^{\int^{\a}_{\a_0}\sqrt{U_p(\a')} d\a'}.
\label{psiII}
\eeq

We now have expressions for solutions $f_p^{I}$ and $f_p^{II}$ in terms of coefficient $B$, and solutions $f_p^{III}$ and $f_p^{IV}$ in terms of coefficient $A$; we must now reconcile solutions everywhere in terms of a single coefficient.  The general solution to the left of $\a_1$ is
\beq
f_p^{II}(\a) = \frac{A'}{2[U_p(\a)]^{1/4}} e^{-\int_{\a}^{\a_1}\sqrt{U_p(\a')} d\a'} + \frac{B'}{[U_p(\a)]^{1/4}} e^{\int_{\a}^{\a_1}\sqrt{U_p(\a')} d\a'}.
\label{II}
\eeq
Matching this across $\a_1$ with the aid of the linearized approximation, we find the form of the solution to the right of $\a_1$,
\beq
f_p^{III}(\alpha) = \frac{e^{-i\pi/4}}{2[-U_p(\alpha)]^{1/4}} \( (A'+iB')e^{i\int_{\alpha_1}^{\alpha}\sqrt{-U_p(\a')} d\a'} + (iA'+B')e^{-i\int_{\a_1}^{\a}\sqrt{-U_p(\a')} d\a'} \).
\eeq

The coefficients $A'$ and $B'$ can now be determined by noting that the solution $f_p^{II}(\a)$ in Eq.~(\ref{II}) must match the solution in Eq.~(\ref{psiII}) determined with the boundary conditions.  Defining
\beq
K = \int_{\alpha_0}^{\alpha_1} \sqrt{U_p(\a')} d\a',
\label{K}
\eeq
we find the relations
\bea
A' &=& 2B e^{K} 
\label{A'}\\
B' &=& \frac{-iB}{2} e^{-K}.
\label{B'}
\eea
Then the solution in region $III$ is
\beq
f_p^{III}(\alpha) = \frac{e^{-i\pi/4}B}{2[-U_p(\alpha)]^{1/4}} \( \(2e^{K}+\frac{e^{-K}}{2}\)e^{i\int_{\alpha_1}^{\alpha}\sqrt{-U_p(\a')} d\a'} + i\(2e^{K}-\frac{e^{-K}}{2}\)e^{-i\int_{\a_1}^{\a}\sqrt{-U_p(\a')} d\a'} \).
\label{f3b}
\eeq

Similarly, we require that the solutions $f_p^{III}(\alpha)$ from Eq.~(\ref{f3a}) and Eq.~(\ref{f3b}) agree:
\bea
& B \( \(2e^{K}+\frac{e^{-K}}{2}\)e^{i\int_{\alpha_1}^{\alpha}\sqrt{-U_p(\a')} d\a'} + i\(2e^{K}-\frac{e^{-K}}{2}\)e^{-i\int_{\a_1}^{\a}\sqrt{-U_p(\a')} d\a'} \) \\
&= A \( e^{i\int_{\alpha}^{\alpha_2}\sqrt{-U_p(\a')} d\a'} + ie^{-i\int_{\a}^{\a_2}\sqrt{-U_p(\a')} d\a'} \).
\eea
Defining
\beq
J = \int_{\alpha_1}^{\alpha_2} \sqrt{-U_p(\a')} d\a' ,
\label{J}
\eeq
the relations
\bea
A&=& iB\(2e^{K}-\frac{e^{-K}}{2}\)e^{-iJ}\\
A&=& -iB\(2e^{K}+\frac{e^{-K}}{2}\)e^{iJ}
\eea
must be simultaneously satisfied.  

In order for a solution to exist, we must require
\beq
\frac{1-\frac{e^{-2K}}{4}}{1+\frac{e^{-2K}}{4}} = -e^{2iJ}
\eeq
or
\beq
J= \(n + \frac{1}{2} \)\pi - \frac{i}{2} \ln\( \frac{1-\frac{e^{-2K}}{4}}{1+\frac{e^{-2K}}{4}} \),
\eeq
where $n$ is an integer.  When $K$ is large, expanding the logarithm results in the approximate relation
\beq
J \simeq \(n + \frac{1}{2} \)\pi + \frac{i}{4} e^{-2K}.
\label{JK}
\eeq
This condition will later be used to determine the momentum eigenvalue $p$ and the decay rate $\Gamma$.

\section{Evaluation of $J$ and $K$}

In the semiclassical regime, assuming that $\gamma$ is not too close to 1 and that $p \ll \beta^{-1}$, the contribution to $J$ from $p$ may be treated as a perturbation:
\bea
J &=& \int_{\alpha_1}^{\alpha_2} \sqrt{-U(\a)+p^2} d\a \\
&\simeq& \int_{\alpha_-}^{\alpha_+} \sqrt{-U(\a)} d\a + \int_{\alpha_-}^{\alpha_+} \frac{p^2}{2\sqrt{-U(\a)}} d\a + \( \int_{\a_1}^{\a_-} \sqrt{-U(\a)} d\a + \int_{\a_+}^{\a_2} \sqrt{-U(\a)} d\a  \) \\
&\equiv& J_0 + J_1 + J_2 .
\eea
We may then evaluate $J_0$, $J_1$, and $J_2$ analytically:
\bea
J_0 &=& \int_{\alpha_-}^{\alpha_+} \sqrt{-U(\a)} d\a = \frac{\pi}{8\beta}(\gamma^3-6\gamma^5+5\gamma^7),
\label{J0}
\\
J_1 &=& \frac{p^2}{2}\int_{\alpha_-}^{\alpha_+} \frac{1}{\sqrt{-U(\a)}} d\a =\frac{\pi \beta p^2}{2},
\\
J_2 &\simeq& \int_{\a_1}^{\a_-} \sqrt{-U(\a)} d\a + \int_{\a_+}^{\a_2} \sqrt{-U(\a)} d\a 
\\
&\simeq& \frac{2}{3} \( \sqrt{-U'(\a_-)} \delta \a_1^{3/2} + \sqrt{U'(\a_+)} \delta \a_2^{3/2} \)\\
&\simeq& \frac{2}{3} \( \frac{p^3}{|U'(\a_-)|} + \frac{p^3}{U'(\a_+)} \) \sim \beta^2 p^3.
\eea
Here, in the calculation of $J_2$ we have expanded $U(\a)$ near $\a_\pm$ and used Eqs.~(\ref{deltaalpha1}), (\ref{deltaalpha2}) and (\ref{U'}).  
The contribution from the correction to the turning points, $J_2$, is small compared to $J_1$; hence we can write 
\beq
J \simeq J_0 + \frac{\pi \beta p^2}{2}.
\label{J2}
\eeq

To evaluate $K$, we again expand about the $p=0$ limit, $K=K_0+\delta K$, where
\beq
K_0 = \int_{-\infty}^{\a_-} \sqrt{U(\a)} d\a
= \frac{1}{24\beta}\[ 15\gamma^6 -13\gamma^4 + \frac{3}{2}\( \gamma^3 - 6\gamma^5 +5\gamma^7 \) \ln\( \frac{\gamma-1}{\gamma+1} \) \].
\label{K2}
\eeq
and $\delta K \sim \beta p^2$ includes all corrections due to $p$.  Even though $\delta K\ll K_0$, we cannot generally neglect $\delta K$.  Since $p\gg 1$ is required for our semiclassical analysis, we can have $\delta K >1$ even if $\beta p\ll 1$. Neglecting $\delta K$ in Eq.~(\ref{JK}) is justified only if 
\beq
\beta p^2 \ll 1,
\eeq
To simplify further analysis, we shall assume this condition to be satisfied.

\section{The decay rate}

We shall now use Eqs.~(\ref{JK}) and (\ref{J2}) to determine the momentum eigenvalue $p$.
We first assign to $p$ real and imaginary parts,
\beq
p=p' + ip'',
\label{p'p''}
\eeq
with $p'$ and $p''$ real.  We shall assume that $p'' \ll p'$; this will be justified below.  (Note that we also neglected the effect of $p''$ on the classical turning points in earlier sections.)

Substituting $J$ from (\ref{J2}) in (\ref{JK}), using (\ref{p'p''}) and neglecting ${p''}^2$ compared to ${p'}^2$, we obtain two relations:
\bea
J_0 + \frac{\pi\beta p'^2}{2}  = \(n + \frac{1}{2} \) \pi \\
\pi\beta p' p'' = \frac{1}{4} e^{-2K_0}.
\label{ImJ}
\eea
With $J_0$ from Eq.~(\ref{J0}), the first of these relations becomes
\beq
\gamma^3(\gamma^2 -1)(5\gamma^2 -1) =4\beta [(2n+1)-\beta p'^2] \approx 4\beta(2n+1).
\label{quantcond}
\eeq
Disregarding the small correction introduced by the ``clock'', as we did in the last step, this is a quantization condition on the parameters of the model $\beta$ and $\gamma$.  Note that if $\gamma$ is not very close to 1, the left hand side of (\ref{quantcond}) is ${\cal O}(1)$, and since $\beta \ll 1$, we must have $n\gg 1$.  The spectrum of the parameters is then nearly continuous, as one would expect in the semiclassical regime. 

The value of $p'$ is largely arbitrary, as long as it satisfies $1\ll p'\ll \beta^{-1/2}$.  Once $p'$ is selected, the imaginary part $p''$ is determined by Eq.~(\ref{ImJ}).  And since $p'\gg 1$ and $\beta\ll 1$ it is easy to see from (\ref{ImJ}) that $p''\ll p'$.

With a complex momentum (\ref{p'p''}), the WDW wave function (\ref{solns}) has the form
\beq
\Psi(\alpha,\phi) = e^{i p'\phi - p'' \phi} f_p(\alpha).
\label{decaysolns}
\eeq
The corresponding probability distribution can be found in terms of the Klein-Gordon current \cite{DeWitt,AV89}.  Up to a normalization constant, it is given by 
\beq
{\cal J}=\frac{i}{2}(\Psi^*\nabla\Psi-\Psi\nabla\Psi^*),
\label{current}
\eeq
In our minisuperspace model, the current has two components,
\bea
{\cal J}^\alpha &=& \frac{i}{2}(\Psi^*\partial_\a \Psi -\Psi\partial_\a \Psi^*), \\ 
{\cal J}^\phi &=& -\frac{i}{2}(\Psi^*\partial_\phi \Psi -\Psi\partial_\phi \Psi^*),
\eea
and satisfies the continuity equation
\beq
\partial_\a{\cal J}^\a +\partial_\phi{\cal J}^\phi = 0.
\eeq
With a proper normalization, the component
\beq
{\cal J}^\phi = p' |f_p(\a)|^2 e^{-2p''\phi}
\eeq
can be interpreted as the probability density for $\a$ at a given "time" $\phi$,
\beq
d{\cal P}\propto {\cal J}^\phi (\a,\phi) d\a .
\eeq 

To express the decay rate in terms of the proper time $t$, we find the amount $\Delta\phi$ by which the field $\phi$ increases during one period of oscillation, $\tau\approx 2\pi/\omega$.  Using the classical equation of motion\footnote{Note that this is different from Eq.~(\ref{pphi}) because of the rescalings $\phi=(4\pi G/3)^{1/2}\varphi$ and $p_\phi = (3/4\pi G)^{1/2} p_\varphi$.} for $\phi$,
\beq
{\dot\phi}=\frac{2G}{3\pi} \frac{p}{a^3}
\eeq
and ignoring, as before, the contribution to the turning points from $p$, we have
\beq
\Delta \phi = \frac{2Gp}{3\pi}\int_\tau dt \frac{1}{a(t)^3} \simeq \frac{4Gp'}{3\pi} \int_{a_-}^{a_+} da \frac{1}{\dot a a(t)^3}.
\eeq
Expressing ${\dot a}$ from Eq.~(\ref{Friedmann}) we evaluate the integral:
\beq
\Delta\phi = 2p' \int_{\a_-}^{\a_+} d\a \frac{1}{\sqrt{-U(\a)}} = 2\pi\beta p'.
\eeq
We now relate the field $\phi$ to the number of oscillations,
\beq
N = \frac{\phi}{\Delta\phi} = \frac{\phi}{2\pi\beta p'},
\eeq
so the probability in Eq.~(\ref{prob}) becomes
\beq
{\cal J}^\phi \propto \ e^{-4\pi\beta p' p'' N} |f_{p'}(\a)|^2.
\label{prob}
\eeq
Finally, using Eq.~(\ref{ImJ}), we obtain
\beq
{\cal J}^\phi \propto \exp\left(e^{-2K_0}N\right).
\label{JKN}
\eeq

We see that the probability for the universe to remain in the oscillating state decreases by a factor of $e$ in $N=e^{2K_0}$ oscillations.  The characteristic lifetime of a simple harmonic universe is thus
\beq
T = \frac{2\pi}{\omega} e^{2K_0} ,
\eeq
with $K_0$ given by Eq.~(\ref{K2}).

\section{The tunneling probability}

In the semiclassical picture, we can think of the SHU as undergoing classical oscillations between the turning points $a_-$ and $a_+$, with some probability of tunneling through the barrier every time it hits the point $a_-$.  We shall now calculate this tunneling probability and relate it to the tunneling rate that we found in the preceding section.

We shall focus on the case of small $\Lambda$, when $\gamma\gg 1$ and the turning points are approximately given by
\bea
a_-\approx \frac{1}{2\gamma\omega} = \frac{3}{8\pi G\sigma}, \\
a_+\approx \frac{2\gamma}{\omega} =\frac{\sigma}{|\Lambda|}.
\eea
The turning points are then widely separated, $a_+/a_- \approx 4\gamma^2 \gg 1$, and the form of the barrier between $a_-$ and $a=0$ is essentially independent of $\Lambda$.  In this regime, we expect the tunneling probability to be nearly the same as for a $\Lambda=0$ universe undergoing a single bounce at $a=a_-$.  Then the probability for SHU to remain in the oscillating phase after $N$ oscillations is
\beq
{\cal P}_N \approx (1-Q)^N \approx e^{-QN},
\label{PN}
\eeq
where $Q\ll 1$ is the tunneling probability for a $\Lambda=0$ universe.

In order to calculate $Q$, we find the semiclassical WDW wave function as we did in Sec. III, except now we only have regions I, II and III to consider.  We do not need a time variable in this case, but the scalar field still plays a useful role of introducing a classically allowed region near the singularity.  This allows us to impose an outgoing boundary condition at $\a\to -\infty$, but we assume as before that the presence of the scalar field has little effect on the dynamics.

By the same argument as in Sec. III, the wave function at large values of $\a$ has the form of (\ref{solns}) with $f_p(\a)$ given by Eq.~(\ref{f3b}).  No boundary condition is imposed at $\a \to +\infty$, so we do not have any quantization condition in this case, and the momentum eigenvalue $p$ can be set to be real.
The wave function (\ref{f3b}) describes an ensemble of contracting universes, which bounce at $\a=\a_-$ and re-expand.  The expanding component has a smaller coefficient, accounting for the fact that some universes have been lost to tunneling decay.  The probability to avoid decay is given by the ratio of the probability fluxes for the two components in Eq.~(\ref{f3b}),
\beq   
1-Q = \frac{{\cal J}^\a(\rightarrow)}{{\cal J}^\a (\leftarrow)} =\left( \frac{4-e^{-2K}}{4+e^{-2K}} \right)^2 \approx 1 - e^{-2K},
\eeq
where left and right arrows correspond to contracting and expanding branches, respectively.  (Note that at large $\a$ both terms in (\ref{f3b}) are very rapidly oscillating, so any interference effects between the two terms become completely negligible.)  Thus, we have
\beq
Q\approx e^{-2K}.
\label{Q}
\eeq

The tunneling exponent $K$ can be found from Eq.~(\ref{K2}).  In the limit of large $\gamma$ it gives
\beq
K_0 \approx \frac{2}{105\beta}\left(1 - \frac{69}{16\gamma^2}\right), 
\eeq
where the second term can be dropped in the limit of $\gamma\to\infty$.

Substituting $Q$ from (\ref{Q}) in Eq.~(\ref{PN}), we recover Eq.~(\ref{JKN}), as expected.

\section{Conclusions}

Our goal in this paper was to implement DeWitt's prescription, that time evolution in quantum cosmology should be described in terms of semiclassical superspace variables, which can be used to define a ``clock''.  We applied this approach to the calculation of the tunneling decay rate of a simple harmonic universe.  The role of a clock in our model was played by a homogeneous, massless, minimally coupled scalar field $\phi$.  The classical evolution of $\phi$ is monotonic, and thus it is a good time variable.

We found the WKB wave functions $\Psi(a,\phi)$, which are eigenstates of the momentum $p_\phi$ conjugate to $\phi$, and matched these wave functions across the turning points, where the WKB approximation breaks down.  We imposed a boundary condition at $a\to\infty$ requiring that $\Psi$ vanishes in that limit and an outgoing boundary condition at $a=0$.  The latter condition means that collapse to $a=0$ is irreversible, so collapsing universes do not bounce back from the singularity.  These two boundary conditions determine the wave function completely and in addition provide two constraints on the parameters of the system and on the momentum eigenvalue $p_\phi$.  We showed how these constraints can be used to calculate the decay rate.  

We also considered the case of a vanishing cosmological constant $\Lambda$, when the universe experiences a single bounce off the barrier and found the tunneling probability through the barrier using the conserved Klein-Gordon-type current.  The resulting probability agrees with our calculation of the decay rate in the limit of small $\Lambda$.

It should not be difficult to extend our analysis to a static universe, which has $\gamma=1$ and $a_+ = a_- =\omega^{-1}$.  In this case, the classically allowed region III reduces to a single point, and the method of a linear approximation for the potential $U(a)$ around the turning points that we used in Sec. III cannot be applied.  However, one can instead use a quadratic approximation $U(a) \propto (a-a_*)^2$  around the point $a_*=\omega^{-1}$.  The wave function in that range can be expressed in terms of harmonic oscillator functions, which will then have to be matched to the WKB wave functions away from $a_*$.  Alternatively, it should be possible to find the solution numerically.

\section*{Acknowledgements}
This work was supported in part by the National Science Foundation under
grant PHY-121388.  A.M. is grateful to Center for Cosmology and Particle 
Physics at New York University for hospitality.

\end{document}